\documentclass[journal]{IEEEtran}

\makeatletter
\let\NAT@parse\undefined
\makeatother
\usepackage[colorlinks=true,linkcolor=black,citecolor=blue,urlcolor=blue,]{hyperref}
\usepackage[doipre={doi:~}]{uri}
\usepackage{amsmath}
\usepackage{latexsym}
\usepackage{algorithmic}
\usepackage{algorithm}

\usepackage{graphicx}
\ifCLASSOPTIONcompsoc 
\usepackage[caption=false,font=normalsize,labelfont=sf,textfont=sf]{subfig}
\else
\usepackage[caption=false,font=footnotesize]{subfig}
\fi
\usepackage{flushend}
\usepackage{xurl}

\begin{document}

\title{A Low-dose CT Reconstruction Network Based on TV-regularized OSEM Algorithm}

\author{Ran~An, Yinghui~Zhang, Xi~Chen, Lemeng~Li, Ke~Chen$^{*}$ and Hongwei~Li$^{*}$

\thanks{Ran An is with the School of Mathematical Sciences, Capital Normal University, Beijing 100048, CHINA, and also with the Centre for Mathematical Imaging Techniques, University of Liverpool, Liverpool L69 7ZL, UK.}

\thanks{Yinghui Zhang is with the LSEC, ICMSEC, Academy of Mathematics and Systems Science, Chinese Academy of Sciences, Beijing 100190, CHINA, and also with the Centre for Mathematical Imaging Techniques, University of Liverpool, Liverpool L69 7ZL, UK.}

\thanks{Xi Chen and Lemeng Li are with the School of Mathematical Sciences, Capital Normal University, Beijing 100048, CHINA.}

\thanks{Ke Chen is with the Department of Mathematics and Statistics, University of Strathclyde, Glasgow G1 1XQ, UK, and also with the Centre for Mathematical Imaging Techniques, University of Liverpool, Liverpool L69 7ZL, UK. (e-mail: k.chen@strath.ac.uk)}

\thanks{Hongwei Li is with the School of Mathematical Sciences, Capital Normal University, Beijing 100048, CHINA. (e-mail: hongwei.li91@cnu.edu.cn)}
}

\markboth{An \MakeLowercase{\textit{et al.}}: A Low-dose CT Reconstruction Network Based on TV-regularized OSEM Algorithm}
{An \MakeLowercase{\textit{et al.}}: A Low-dose CT Reconstruction Network Based on TV-regularized OSEM Algorithm}

\maketitle

\begin{abstract}
Low-dose computed tomography (LDCT) offers significant advantages in reducing the potential harm to human bodies. However, reducing the X-ray dose in CT scanning often leads to severe noise and artifacts in the reconstructed images, which might adversely affect diagnosis. By utilizing the expectation maximization (EM) algorithm, statistical priors could be combined with artificial priors to improve LDCT reconstruction quality. However, conventional EM-based regularization methods adopt an alternating solving strategy, i.e. full reconstruction followed by image-regularization, resulting in over-smoothing and slow convergence. In this paper, we propose to integrate TV regularization into the ``M''-step of the EM algorithm, thus achieving effective and efficient regularization. Besides, by employing the Chambolle-Pock (CP) algorithm and the ordered subset (OS) strategy, we propose the OSEM-CP algorithm for LDCT reconstruction, in which both reconstruction and regularization are conducted view-by-view. Furthermore, by unrolling OSEM-CP, we propose an end-to-end reconstruction neural network (NN), named OSEM-CPNN, with remarkable performance and efficiency that achieves high-quality reconstructions in just one full-view iteration. Experiments on different models and datasets demonstrate our methods' outstanding performance compared to traditional and state-of-the-art deep-learning methods.
\end{abstract}


\section{Introduction}
CT is a widely used technique in medical diagnosis. It is known that excessive X-ray radiation could have potential harm to the human body, including genetic damage and cancer risk \cite{lidestaahl2021estimated}. Therefore, the as low as reasonably achievable (ALARA) principle \cite{ALARA_solomon2020justification} was proposed to guide the dose of X-rays in CT diagnosis. Hence, LDCT reconstruction has always been a popular subject in the medical imaging community. 

In LDCT reconstruction, traditional algorithms such as the filtered back-projection (FBP) \cite{FBP_willemink2019evolution} and simultaneous algebraic reconstruction technique (SART) \cite{SART_andersen1984simultaneous} usually introduce severe noise and artifacts in the reconstructed image \cite{LDCT_goldman2007principles, pan2009commercial}. To obtain high-quality reconstructions, some denoising strategies are introduced for projection data pre-processing \cite{sinopre3_manduca2009projection, sinopre4_zhang2010statistical, sinopre5_balda2012ray} and image post-processing \cite{imgpost1_chen2009bayesian, imgpost2_kang2013image, imgpost3_green2016efficient}. However, these methods often bring excessive blurring or secondary artifacts in the reconstructed images. Some other works integrate artificial priors such as the total variation (TV) \cite{TV_rudin1994total} into the reconstruction procedure and propose regularized iterative algorithms which could demonstrate superior reconstructions \cite{SARTTV_yu2009compressed, epTV_tian2011low, sidky2011constrained}. To overcome the slow convergence issue of TV-regularized iterative algorithms, Deng \textit{et al.} incorporated TV regularization into the OS-SART and solved the reconstruction model with the CP algorithm \cite{CP_chambolle2011first}, thus proposing the OS-CP \cite{OSCP_deng2019fast} algorithm. Nevertheless, the computational burden is still high. In addition, the tuning work for the hyper-parameters involved in regularization models could be tedious. 

Recently, deep learning (DL) has attracted wide attention with its powerful ability in model representation and data fitting. In LDCT reconstruction, DL methods have shown great success. Among them, DL-based image post-processing is a commonly used strategy, with representative methods including the FBPConvNet \cite{FBPConvNet_jin2017deep}, RED-CNN \cite{REDCNN_chen2017low}, DIRE \cite{DIRE_liu2019deep} and CTFormer \cite{wang2023ctformer}. They train denoising NN with paired normal-dose and low-dose CT images, achieving remarkable improvement compared with traditional methods. However, these purely post-processing methods can easily cause over-blurring and detail loss in the reconstructed images. Therefore, some dual-domain methods were proposed that integrate pre-processing and post-processing in a whole network, such as the DRCNN \cite{drcnn_feng2021dual}, DDPNet \cite{DDPNet_ge2022ddpnet}, DuDoUFNet \cite{zhou2022dudoufnet} and DRONE \cite{wu2021drone}. Although with powerful performance, dual-domain methods still suffer problems. Denoising in the sinogram domain is a delicate problem and usually causes secondary artifacts if not done properly. In addition, the commonly used mean squared error (MSE) loss often leads to blurring and structure distortions due to the averaging trend. To enhance the visual effects, some methods based on the generative adversarial networks (GAN) \cite{GAN_creswell2018generative} have been proposed to generate clear and sharp images, such as the CaGAN \cite{cagan_huang2020cagan} and SAGAN \cite{sagan_yi2018sharpness}. Some other GAN-based methods enhance the image results after NN denoising, such as the DD-UNET \cite{ddunet_zheng2020dual} and CLEAR \cite{clear_zhang2021clear}. GAN has substantial advantages in enhancing the details and visual effects of images. However, ensuring the accuracy of the generated images is challenging. In addition, GAN is prone to model collapse due to bad training stability, which would also bring troubles.

On the other hand, some methods use NN to learn the priors in data and integrate them into the iterative reconstruction algorithms. The deep unrolling approaches such as the LEARN \cite{learn_chen2018learn} and PD-Net \cite{learnpd_adler2018learned} regroup iterative algorithms using NN to replace the involved operators and hyperparameters. Through adaptive training, they achieve better reconstructions with even fewer training samples. Some other methods utilize NN as regularization terms. Wu \textit{et al.} employed an unsupervised K-sparse auto-encoder to extract image priors \cite{wu2017iterative}. Fabian \textit{et al.} trained normalizing flows PatchNR \cite{PatchNR_altekruger2023patchnr} to learn priors in patches of the normal-dose images and combine it as a regularization term in iterative reconstructions. Similarly, He \textit{et al.} and Liu \textit{et al.} proposed the EASEL \cite{EASEL_he2022iterative} and Dn-Dp \cite{DNDP_liu2023diffusion} respectively, based on the regularization terms learned by the diffusion model \cite{Diffusion_croitoru2023diffusion}. Despite their outstanding performance, these methods still have drawbacks. They often require multiple iterations and large computations as traditional iterative algorithms do. 

Statistical priors can also help improve LDCT reconstructions. With a suitable prior distribution, the EM \cite{EM_moon1996expectation} algorithm can be employed for iterative reconstruction. In nuclear imaging, Sheep and Vardi first introduced the EM algorithm into emission tomography \cite{shepp1982maximum}, then Lange and Carson used the EM algorithm in transmission tomography \cite{lange1984reconstruction}. Although showing great success in emission tomography such as for PET and SPECT \cite{lange1984reconstruction} imaging, the EM algorithm encounters problems when applied in transmission tomography CT imaging. It's known that the transmission EM algorithm does not have a closed-form solution \cite{celler1998multiple} in the ``M''-step. Although some approximations exist, either the statistical priors might not be well preserved, or the calculations might be unstable \cite{ollingery1992use,lange1987theoretical}. The EM algorithm assumes that the log-transformed projection data obey the Poisson distribution \cite{imgpost1_chen2009bayesian, sheng2018adjustable, chen2020low, friot2022iterative}. However, such an assumption does not match the physical truth well and this model bias would lead to image degradation such as noise reflux \cite{dempster1977maximum}. Although some works represented by the MLEM-TV \cite{mlemtv_chavez2015ml} introduce regularization to guide the statistical reconstruction process \cite{mlemtv_chavez2015ml, mlemnlm_zhang2014nonlocal, mlemcdr_kaur2019low}, they mostly use an alternating strategy between EM reconstruction and image regularization, thus the regularization does not really constrain the reconstruction procedure, and over-blurring and detail loss might still happen.

In this paper, we propose a novel statistical LDCT reconstruction framework. For the TV-regularized EM reconstruction, instead of the alternating solving strategy, we combine the regularization into the ``M''-step of the EM reconstruction process and use the CP algorithm \cite{CP_chambolle2011first} to solve the minimization problem, thus effectively integrating the statistical priors into the reconstruction process. Besides, we introduce the OS strategy \cite{osem_hudson1994accelerated} into the iterative reconstruction to conduct view-by-view reconstruction and regularization, thus the reconstruction is greatly accelerated. We name our proposed TV-regularized statistical iterative algorithm as \textbf{OSEM-CP}.

Furthermore, we propose to unroll our OSEM-CP algorithm into an end-to-end reconstruction neural network to leverage the priors in ``big data'', thus developing \textbf{OSEM-CPNN}. With adaptive learning of hyper-parameters and proximal operators, OSEM-CPNN significantly improves the reconstruction quality by just one full iteration while eliminating the tuning work for artificial hyperparameters. Our work can be summarized as follows:
\begin{itemize}
 \item We propose a novel TV-regularized EM algorithm OSEM-CP for LDCT reconstruction. With the help of the CP algorithm and the OS strategy, statistical priors are effectively and efficiently integrated into the reconstruction procedure.
 \item We propose an end-to-end LDCT reconstruction neural network OSEM-CPNN by unrolling the proposed iterative reconstruction algorithm OSEM-CP. While removing the hyperparameter-tuning workload, the proposed network significantly improves the reconstruction quality with just one full iteration (all views projection data are processed only once). 
 \item Experimental results demonstrate that our methods achieve promising performance in LDCT reconstruction. Our OSEM-CP outperforms popular traditional methods, while the proposed OSEM-CPNN surpasses state-of-the-art supervised NN methods.
\end{itemize}

\section{Related Work}
\subsection{EM Algorithm for Emission Tomography}
The EM algorithm for emission tomography is based on the projection data obeying a Poisson distribution:
\begin{eqnarray}
	\mathbf{p} \sim \mathrm{Poisson}\left( \mathbf{A}\mathbf{x} \right),
	\label{eq_2.1}
\end{eqnarray}
where $\mathbf{p}=[p_{1}, p_{2}, ..., p_{M}]^{T}$ represents the projection data vector, $\mathbf{x}=[x_{1}, x_{2}, ..., x_{N}]^{T}$ represents the vectorized image and $\mathbf{A}=[a_{ij}]_{M\times N}$ is the projection matrix. Based on this assumption, we can introduce a Poisson distributed latent vector variable $\mathbf{c}=\{c_{ij}:i=0,1,...,M;j=0,1,...,N\}$ where $c_{ij}$ represents the contribution of the $j_{th}$ pixel $\mathbf{x}_{j}$ to the $i_{th}$ item of the projection data $p_{i}$, that is to say, $p_{i}=\sum^N_{j=1} {{c_{ij}}}$. And the expectation $\eta_{ij}$ of $c_{ij}$ is:
\begin{eqnarray}
	\eta_{ij} = E\left(c_{ij}\right) = a_{ij}{x_j}.
	\label{eq_2.2}
\end{eqnarray}
Then the log-likelihood function about $\mathbf{c}$ and the image $\mathbf{x}$ can be expressed as:
\begin{eqnarray}
    \ln P\left(\mathbf{c}|\mathbf{x}\right) = \sum\limits_{i,j}(c_{ij}\ln(a_{ij}x_j) - a_{ij}x_j) - \sum\limits_{i,j}\ln(c_{ij}!).
	\label{eq_2.3}
\end{eqnarray}
The last term on the right side of the equation (\ref{eq_2.3}) is constant, thus solving $\mathbf{x}$ is equivalent to finding the maximum of the following objective function:
\begin{eqnarray}
	L = \sum\limits_{i,j} {\left( {{c_{ij}}\ln \left( {{a_{ij}}{x_j}} \right) - {a_{ij}}{x_j}} \right)}. 
	\label{eq_2.4}
\end{eqnarray}

There are two steps for solving the maximum likelihood problem (\ref{eq_2.4}) with the EM algorithm: the expectation (\textit{``E''}) step and the maximization (\textit{``M''}) step. At the expectation step of the $(n+1)_{th}$ iteration, the latent variable $c_{ij}$ is replaced with its current expectation:
\begin{eqnarray}
	E\left( {{c_{ij}}\left| {\mathbf{p},{\mathbf{x}^n}} \right.} \right) = \frac{{{a_{ij}}x_j^n}}{{\sum\limits_k {{a_{ik}}x_k^n} }}{p_i},
	\label{eq_2.5}
\end{eqnarray}
where $\mathbf{x}^n$ is the image in the $n_{th}$ iteration and $x_j^n$ is the $j_{th}$ pixel of $\mathbf{x}^n$. Thus, the objective function (\ref{eq_2.4}) can be expressed as an expectation containing the projection data $\mathbf{p}$:
\begin{eqnarray}
	E\left( {L\left|\mathbf{p},\mathbf{x}^n\right.} \right) = \sum\limits_{i,j} {\left( {\frac{{{a_{ij}}x_j^n}}{{\sum\limits_k {{a_{ik}}x_k^n} }}{p_i}\ln \left( {{a_{ij}}{x_j}} \right) - {a_{ij}}{x_j}} \right)}. 
	\label{eq_2.6}
\end{eqnarray}
At the maximization step, by maximizing the expectation (\ref{eq_2.6}), the solution of $x_j$ in the $(n+1)_{th}$ iteration is given by:
\begin{eqnarray}
	x_j^{n + 1} = \frac{{x_j^n}}{{\sum\limits_i {{a_{ij}}} }}\sum\limits_i {{a_{ij}}\frac{{{p_i}}}{{\sum\limits_k {{a_{ik}}x_k^n}}}}.
	\label{eq_2.7}
\end{eqnarray}

\subsection{EM Algorithm for Transmission Tomography}
The prior distribution of the transmission EM model is that the number of photons passing through each pixel follows a Poisson distribution about the number of incident photons and the photon attenuation \cite{lange1984reconstruction}. Introducing a latent vector variable $\mathbf{I}=\{I_{ij}:i= 0,1,...,M;j=0,1,...,N\}$ where $I_{ij}$  represents the photon number of the $i_{th}$ ray passing through the $j_{th}$ pixel, its distribution can be expressed as:
\begin{equation}
    {I_{ij}}\sim \mathrm{Poisson}\left(I_i \prod \limits_{l=1}^j e^{-a_{il}x_l} \right),
    \label{eq_2.8}
\end{equation}
where $I_i$ is the incident photon number of the $i_{th}$ ray. The log-likelihood function then is given by:
\begin{eqnarray}
    \ln P\left(\mathbf{I}|\mathbf{x}\right) = \sum\limits_{i,j} P(I_{i})\ln P(I_{ij}|I_{i,j-1};\mathbf{x}).
	\label{eq_2.9}
\end{eqnarray}
After introducing the prior distribution and simplifying the constant terms, the objective function to maximize is:
\begin{equation}
    L = \sum\limits_{i,j} (I_{ij}\ln(e^{-a_{ij}{x_j}}) + (I_{i,j-1}-I_{ij})\ln(1-e^{-a_{ij}{x_j}})).
    \label{eq_2.10}
\end{equation}

At the expectation step of the $(n+1)_{th}$ iteration, the latent variable $I_{ij}$ and $I_{i,j-1}$ are substituted with their expectations $N_{ij}^n$ and $M_{ij}^n$, which gives:
\begin{equation}
\begin{aligned} 
    E(L|\mathbf{I},\mathbf{x}^n) = &\sum\limits_{i,j} {\bar N}^n_{ij}\ln(e^{-a_{ij}{x_j}}) + \\
    &\sum\limits_{i,j} ({\bar M}^n_{ij}-{\bar N}^n_{ij})\ln(1-e^{-a_{ij}{x_j}}),
    \label{eq_2.11}
\end{aligned}
\end{equation}
where
\begin{equation}
\begin{aligned}
    {\bar N}^n_{ij} &= E(I_{ij}|{I_i},\mathbf{x}^n), \\
    {\bar M}^n_{ij} &= E(I_{i,j-1}|{I_i},\mathbf{x}^n).
    \label{eq_2.12}
\end{aligned}
\end{equation}
At the maximization step, no closed-form solution can be derived by maximizing the $E(L|\mathbf{I},\mathbf{x}^n)$ in equation (\ref{eq_2.11}). Via Taylor expansion, Lange and Carson gave an approximate iterative scheme for solving (\ref{eq_2.11}) \cite{lange1984reconstruction}:
\begin{equation}
    x_j^{n+1} \approx \frac{\sum\limits_i( {\bar M}^n_{ij}-{\bar N}^n_{ij})}{\frac{1}{2}\sum\limits_i({\bar M}^n_{ij}+{\bar N}^n_{ij})}.
    \label{eq_2.13}
\end{equation}
This iterative scheme leads to a higher computational burden and possible loss of statistical priors. 

\subsection{OSEM Algorithm}
The EM algorithm shows good respect for the statistical properties of the projection data, and thus fits well to LDCT reconstruction. However, as an iterative reconstruction method, it still has the disadvantage of a heavy computational burden compared to the analytical methods. To improve the efficiency of the EM algorithm, Hudson et al. proposed the OSEM algorithm \cite{osem_hudson1994accelerated}. With an OS technique, the projection data are sorted and divided into $M$ ordered subsets. The EM algorithm is then conducted within these subsets one by one, and one complete iteration traverses all $M$ subsets. Mathematically, similar to equation (\ref{eq_2.7}), the OSEM algorithm for emission tomography is given by:
\begin{eqnarray}
	x_j^{n + 1} = \frac{{x_j^n}}{{\sum\limits_{i \in {S_m}} {{a_{ij}}} }}\sum\limits_{i \in {S_m}} {{a_{ij}}\frac{{{p_i}}}{{\sum\limits_k {{a_{ik}}x_k^n} }}},
	\label{eq_2.14}
\end{eqnarray}
where ${S_m}$ is the $m_{th}$ subset, $m = 1,2,...,M$.
\section{Methods}
\subsection{OSEM-CP Algorithm}
The CT scanning is a typical transmission process, thus ideally the transmission EM algorithm should be applied for LDCT reconstruction. Unfortunately, as stated previously, the transmission EM algorithm is accompanied by high computational complexity and possible loss of statistical priors due to the approximate solver in its ``M''-step. Considering that both emission and transmission EM algorithms could be regarded as solving linear systems like $\mathbf{A}\mathbf{x}=\mathbf{p}$, we can still use the emission EM algorithm for CT reconstruction at the price of partially losing statistical priors, since the emission and transmission processes obey different statistical processes. Indeed, when the forward process is modeled as a linear system, its statistical properties, e.g. noise distribution, are simply dropped out. This mismatch of statistical priors might adversely affect the reconstructed image. Introducing artificial regularization is an effective way to remediate model mismatching, however, existing regularized methods such as MLEM-TV \cite{mlemtv_chavez2015ml} use an alternating updating strategy between full reconstruction and image regularization, so the regularization is not tightly combined with the EM reconstruction procedure, easily causing excessive smoothing and loss of image details. In addition, such an alternating strategy usually suffers from slow convergence.

In this paper, we propose an effective way to integrate artificial priors into the EM reconstruction process, and by adopting the OS strategy and the CP \cite{CP_chambolle2011first} algorithm, both the regularization and reconstruction are conducted in a view-by-view way, which leads to a fast and stable algorithm for LDCT reconstruction. Our method has better image-data consistency and maintains the non-negativity property of the EM algorithm. We would like to name our method as \textbf{OSEM-CP}, which is explained in detail below.

At the maximization step, i.e. ``M''-step, of the emission EM algorithm, the maximization problem for the expectation (\ref{eq_2.6}) can be rewritten as minimizing the following objective function:
\begin{equation}
	L(\mathbf{x};{\mathbf{x}^n}) 
 = \sum\limits_{i,j} {\left( {{a_{i,j}}{x_j} - \frac{{{a_{i,j}}x_j^n}}{{\sum\limits_k {{a_{i,k}}x_k^n} }}{p_i}\ln \left( {{a_{i,j}}{x_j}} \right)} \right)}. 
	\label{eq_3.15}
\end{equation}

Introducing the TV-regularization term $R(x)=\|\nabla\mathbf{x}\|_1$ as a constraint, this problem becomes a regularized minimization problem with the objective function:
\begin{equation}
\begin{aligned}
	H&\left( {\mathbf{x};{\mathbf{x}^n}} \right) = L(\mathbf{x};{\mathbf{x}^n}) + \lambda R(\mathbf{x}) \\
 &= \sum\limits_{i,j} {\left( {{a_{i,j}}{x_j} - \frac{{{a_{i,j}}x_j^n}}{{\sum\limits_k {{a_{i,k}}x_k^n} }}{p_i}\ln \left( {{a_{i,j}}{x_j}} \right)} \right)} + \lambda {\left\| {\nabla \mathbf{x}} \right\|_1},
	\label{eq_3.16}
\end{aligned}
\end{equation}
where $\lambda$ is a manual hyper-parameter to control the regularization effects, and 
\begin{equation}
	\|\nabla\mathbf{x}\|_1 = \sum^N\limits_{i,j=1}{|(\nabla \mathbf{x})}_{i,j}|
	\label{eq_3.17}
\end{equation}
is the isotropic discrete total variation of an image $\mathbf{x} \in \mathbf{R}^{N\times N}$.

To minimize the objective function $H$, we employ the CP algorithm \cite{CP_chambolle2011first} which is a first-order primal-dual algorithm commonly used for solving convex problems of the following form:
\begin{equation}
	\mathop {\min }\limits_{\mathbf{x}} \left\{ {F\left( {\mathrm{K}\mathbf{x}} \right) + G\left( \mathbf{x} \right)} \right\},
	\label{eq_3.18}
\end{equation}
where $F$ and $G$ are two convex functions, $\mathrm{K}$ is a continuous linear operator. The convex conjugate of $F(\mathrm{K}\mathbf{x})$ can be expressed as:
\begin{equation}
	F(\mathrm{K}\mathbf{x}) = \mathop {\max }\limits_{\mathbf{y}} \left\{ {\left\langle {\mathrm{K}\mathbf{x},\mathbf{y}} \right\rangle - {F^*}\left( \mathbf{y} \right)} \right\},
	\label{eq_3.19}
\end{equation}
where $F^*$ is the dual form of $F$. The prime-dual form of problem (\ref{eq_3.18}) then can be derived as:
\begin{equation}
	\mathop {\min }\limits_{\mathbf{x}} \mathop {\max }\limits_{\mathbf{y}} \left\langle {\mathrm{K}\mathbf{x},\mathbf{y}} \right\rangle + G\left( \mathbf{x} \right) - {F^*}\left( \mathbf{y} \right).
	\label{eq_3.20}
\end{equation}
With the CP algorithm, the $R(\mathbf{x})$ and $L(\mathbf{x};{\mathbf{x}^n})$ in the objective function (\ref{eq_3.16}) can be expressed as:
\begin{align}
 &F(\nabla \mathbf{x}) = \lambda \| \nabla \mathbf{x} \|_{1}, \label{eq_3.21}\\
 &G(\mathbf{x}) = \sum\limits_{i,j} \left( a_{i,j}x_{j}-\frac{a_{i,j}x_{j}^{n}}{\sum\limits_{k}{a_{i,k}x_{k}^{n}}}p_{i}\ln(a_{i,j}x_{j})\right). \label{eq_3.22}
\end{align}

Introducing an auxiliary variable $\mathbf{q}$, the prime-dual form of the regularized minimization problem is given by:
\begin{equation}
\begin{aligned}
	\mathop {\min }\limits_{\mathbf{x}} \mathop {\max }\limits_{\mathbf{q}} \langle \nabla_\lambda \mathbf{x},\mathbf{q} \rangle &+ \sum\limits_{i,j} \left( a_{i,j}x_{j}-\frac{a_{i,j}x_{j}^{n}}{\sum\limits_{k}{a_{i,k}x_{k}^{n}}}p_{i}\ln(a_{i,j}x_{j}) \right) \\
 &-\delta_\mathbf{q}(\mathbf{q}).
	\label{eq_3.23}
\end{aligned}
\end{equation}
where $\delta_\mathbf{q}(\mathbf{q})$ is the dual form of the TV-regularization term (\ref{eq_3.17}), in detail:
\begin{equation}
\begin{aligned} 
\delta_\mathbf{q}(\mathbf{q}) = \left\{ 
\begin{array}{cr} 
 0 &\mathbf{q} \in Q, \\ 
 \infty &\mathbf{q} \notin Q,  
\end{array} 
\right. 
\label{eq_3.24}
\end{aligned}
\end{equation}
where $Q = \left\{ \mathbf{q}: \left\| {\nabla \mathbf{q}} \right\|_\infty \le 1 \right\}$. With the proximal gradient algorithm, we can solve $\mathbf{x}$ and $\mathbf{q}$ in equation (\ref{eq_3.23}) iteratively:
\begin{equation}
\left\{
\begin{array}{cr}
\begin{aligned}
	\mathbf{q}^{n+1} &= \mathrm{prox}_{{\sigma}{\delta_\mathbf{q}}}\left( {{\mathbf{q}^n} + {\sigma}{\nabla _\lambda}{{\bar{\mathbf{x}}}^n}} \right),\\ 
 {\mathbf{x}^{n + 1}} &= \mathrm{prox}_{{\tau}f}\left( {{\mathbf{x}^n} + {\tau}\mathrm{div}_\lambda {\rm{ }}{\mathbf{q}^{n + 1}}} \right), \\ 
 {\bar{\mathbf{x}}^{n + 1}} &= 2{\mathbf{x}^{n + 1}} - {\mathbf{x}^n},
\end{aligned}
\end{array} 
\right.
\label{eq_3.25}
\end{equation}
where $\sigma$ and $\tau$ are two hyperparameters representing the step sizes and $f$ is the objective function on a single pixel. Let $\tilde{\mathbf{q}} = \mathbf{q}^n + {\sigma}{\nabla _\lambda }{\bar{\mathbf{x}}^n}$ and $\tilde{\mathbf{x}} = {\mathbf{x}^n} + {\tau}di{v_\lambda }{\rm{ }}{\mathbf{q}^{n + 1}}$, we have:
\begin{equation}
	{\mathbf{q}^{n + 1}} = \mathrm{prox}_{{\sigma}{\delta_\mathbf{q}}}\left( {\tilde{\mathbf{q}}} \right) \Leftrightarrow q_{i.j}^{n + 1} = \frac{{{{\tilde q}_{i,j}}}}{{\max \left( {1,\left| {{{\tilde q}_{i,j}}} \right|} \right)}},
\label{eq_3.26}
\end{equation}
and
\begin{equation}
\begin{aligned}
 x_j^{n + 1} &= \mathrm{prox}_{{\tau}f}\left( {{{\tilde x}_j}} \right)\\
 &= \mathop {\arg \min }\limits_{u_j} \left\{ {{\tau}f\left( {{u_j}} \right) + \frac{1}{2}{{\left\| {{u_j} - {{\tilde x}_j}} \right\|}^2}} \right\}.
	\label{eq_3.27}
\end{aligned}
\end{equation}
Solving the optimization problem (\ref{eq_3.27}) we have:
\begin{equation}
\begin{aligned}
		&{\tau}\sum\limits_i {\left( {{a_{i,j}} - \frac{{{a_{i,j}}x_j^n}}{{\sum\limits_k {{a_{i,k}}x_k^n} }}{p_i}\frac{{{a_{i,j}}}}{{{a_{i,j}}{u_j}}}} \right)} + \left( {{u_j} - {{\tilde x}_j}} \right) = 0 \\
  \Leftrightarrow \rm{ } &{\tau}\frac{1}{{{u_j}}}\sum\limits_i {\left( {{a_{i,j}}{u_j} - \frac{{{a_{i,j}}x_j^n}}{{\sum\limits_k {{a_{i,k}}x_k^n} }}{p_i}} \right)} + \left( {{u_j} - {{\tilde x}_j}} \right) = 0 \\ 
  \Leftrightarrow \rm{ } &u_j^2 + {u_j}\left( {{\tau}\sum\limits_i {\left( {{a_{i,j}}} \right) - } {{\tilde x}_j}} \right) - {\tau}x_j^n\sum\limits_i {{a_{i,j}}} \frac{{{p_i}}}{{\sum\limits_k {{a_{i,k}}x_k^n} }} \\
  &\quad = 0.
	\label{eq_3.28}
\end{aligned}
\end{equation}
Equation (\ref{eq_3.28}) is a problem of solving a quadratic equation from which the solution can be derived as:
\begin{figure*}[!t]
	\centering
	\includegraphics[width=7.1in]{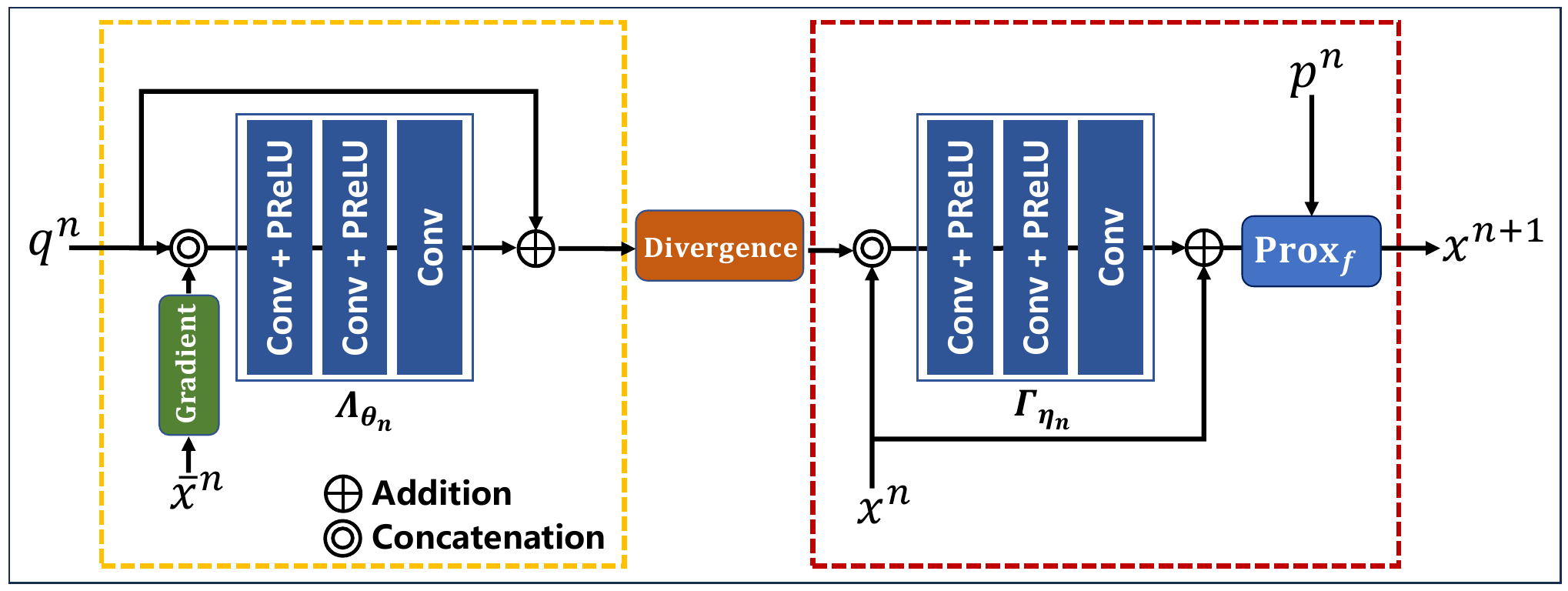}
  \vspace{-0.3cm}
	\caption{The structure of the $n_{th}$ sub-module in the proposed OSEM-CPNN.}
	\label{fig_1_net}
\end{figure*}
\begin{equation}
\begin{aligned}
	&x_j^{n + 1} = -\frac{1}{2} \left( {{\tau}\sum\limits_i {\left( {{a_{i,j}}} \right) - } {{\tilde x}_j}} \right) \pm \\ 
 &\frac{1}{2}\sqrt{{{\left( {{\tau}\sum\limits_i {\left( {{a_{i,j}}} \right) - } {{\tilde x}_j}} \right)}^2} + 4{\tau}x_j^n\sum\limits_i {{a_{i,j}}} \frac{{{p_i}}}{{\sum\limits_k {{a_{i,k}}x_k^n} }}}.
	\label{eq_3.29}
\end{aligned}
\end{equation}

Considering the non-negativity of the image pixel values, the positive solution is chosen as the update of $\mathbf{x}$, which guarantees the non-negativity of the solution as the EM algorithm does. 

To accelerate the convergence of our method, we utilize the OS strategy in the iterative reconstruction. In detail, the order of projection angles is scrambled and each subset contains the projection data corresponding to one projection view (angle). Thus we conduct a view-by-view reconstruction and the regularization will also be applied view-by-view. The proposed OSEM-CP algorithm is summarized in \textbf{Algorithm 1}.

\begin{algorithm}
	\caption{OSEM-CP}
	\begin{algorithmic}[1]
		\STATE The projection data $\mathbf{p}$, the sequence of ordered subsets $1,...,M$, the number of iterations $T$.
		\STATE Setting the initial value $\mathbf{x}^{0} > \mathbf{0}$, $\mathbf{q}^{0} > \mathbf{0}$, the iteration step size $\tau>0, \sigma>0$, the hyper-parameter for the TV-regularization term $\lambda >0$.
		\STATE for $n \le T$:
		\STATE \quad for $m < M$:
		\STATE \quad \quad ${\mathbf{q}}^{n+\frac{m+1}{M}}\leftarrow \mathrm{prox}_{\sigma {\delta_{\mathbf{q}}}}\left( {\mathbf{q}^{n+\frac{m}{M}}}+\sigma {{\nabla }_{\lambda }}{{{\bar{\mathbf{x}}}}^{n+\frac{m}{M}}} \right)$.
		\STATE \quad \quad ${\mathbf{x}^{n+\frac{m+1}{M}}}\leftarrow \mathrm{prox}_{\tau f}\left( {\mathbf{x}^{n+\frac{m}{M}}}+\tau \mathrm{div}_{\lambda }{\mathbf{q}^{n+\frac{m+1}{M}}} \right)$.
		\STATE \quad \quad ${{\bar{\mathbf{x}}}^{n+\frac{m+1}{M}}}\leftarrow 2{\mathbf{x}^{n+\frac{m+1}{M}}}-{\mathbf{x}^{n+\frac{m}{M}}}$.
		\STATE return $\mathbf{x}^{T}$
	\end{algorithmic}
\end{algorithm}

\begin{algorithm}
	\caption{OSEM-CPNN}
	\begin{algorithmic}[1]
		\STATE The projection data $\mathbf{p}$, the sequence of ordered subsets $1,...,M$.
		\STATE Setting the initial value $\mathbf{x}^{0} > \mathbf{0}$, $\mathbf{q}^{0} > \mathbf{0}$.
		\STATE for $m < M$:
		\STATE \quad ${\mathbf{q}^{\frac{m+1}{M}}}\leftarrow \Lambda_{\theta_m}\left( {\mathbf{q}^{\frac{m}{M}}},\nabla \bar{\mathbf{x}}^{\frac{m}{M}} \right)$.
		\STATE \quad $\tilde{\mathbf{x}}^{\frac{m+1}{M}}\leftarrow \Gamma_{\eta_m}\left( {\mathbf{x}^{\frac{m}{M}}},\mathrm{div}{\mathbf{q}^{\frac{m+1}{M}}} \right)$. 
		\STATE \quad ${\mathbf{x}{^\frac{m+1}{M}}}=\mathrm{prox}_{f}\left( {\tilde{\mathbf{x}}{^\frac{m+1}{M}}} \right)$.
		\STATE \quad ${{\bar{\mathbf{x}}^\frac{m+1}{M}}}=2{\mathbf{x}^\frac{m+1}{M}}-\mathbf{x}^\frac{m}{M}$.
		\STATE return $\mathbf{x}^1$
	\end{algorithmic}
\end{algorithm}

\subsection{OSEM-CPNN}
In our OSEM-CP algorithm, the TV-regularization term and CP algorithm introduce hyperparameters $\lambda$, $\tau$, and $\sigma$, which require manual adjustments and thus increase the workload for performance tuning. In addition, the explicit proximal operators are approximations and might not be the best choices. In light of deep neural networks, learning the hyperparameters and proximal operators through a data-driven approach is a reasonable and efficient strategy. By unrolling the iterations into neural network layers and adaptively learning the hyperparameters and proximal operators, the performance and efficiency of the reconstruction shall be significantly improved. Successful cases have been demonstrated, for example, in \cite{learn_chen2018learn,learnpd_adler2018learned}.

So, to utilize data-driven priors, we unrolled our OSEM-CP algorithm into an end-to-end reconstruction network, named \textbf{OSEM-CPNN}. By eliminating the workload for tuning hyperparameters and reducing the number of full-view iterations of OSEM-CP to only once, i.e. set $T=1$ in \textbf{Algorithm 1}, OSEM-CPNN significantly improves the performance and efficiency of LDCT reconstruction, compared to OSEM-CP and other popular LDCT reconstruction methods. For LDCT reconstruction, traditional iterative methods need multiple full-view iterations to achieve reasonable reconstructions, and a faithful unrolling of OSEM-CP would lead to $T>1$. According to our experiments, however, setting $T>1$ earns marginal improvements.

In each iteration of the OSEM-CPNN, which performs reconstruction with the projection data of a specific view, we use a convolutional neural network (CNN) $\Lambda_\theta$ to replace the proximal operator $\mathrm{prox}_{\sigma \delta_{q}}$ of the dual problem. For the prime problem, considering that the proximal operator is complex and the solution helps to guarantee the non-negativity property, we retain it and learn the hyperparameters with another CNN $\Gamma_\eta$. The hyperparameters $\sigma$ and $\lambda$ of the original algorithm are integrated into the CNNs, and the parameter $\tau$ in (\ref{eq_3.26}) is set as a learnable parameter. As a noteworthy advantage, OSEM-CPNN only requires one full-view (angle) iteration to produce high-quality images. The whole network is trained by the \textbf{MSE} loss between the output images and the accessible labels. The main steps of OSEM-CPNN are outlined in \textbf{Algorithm 2}.

\begin{figure*}[!t]
	\centering
	\includegraphics[width=7.1in]{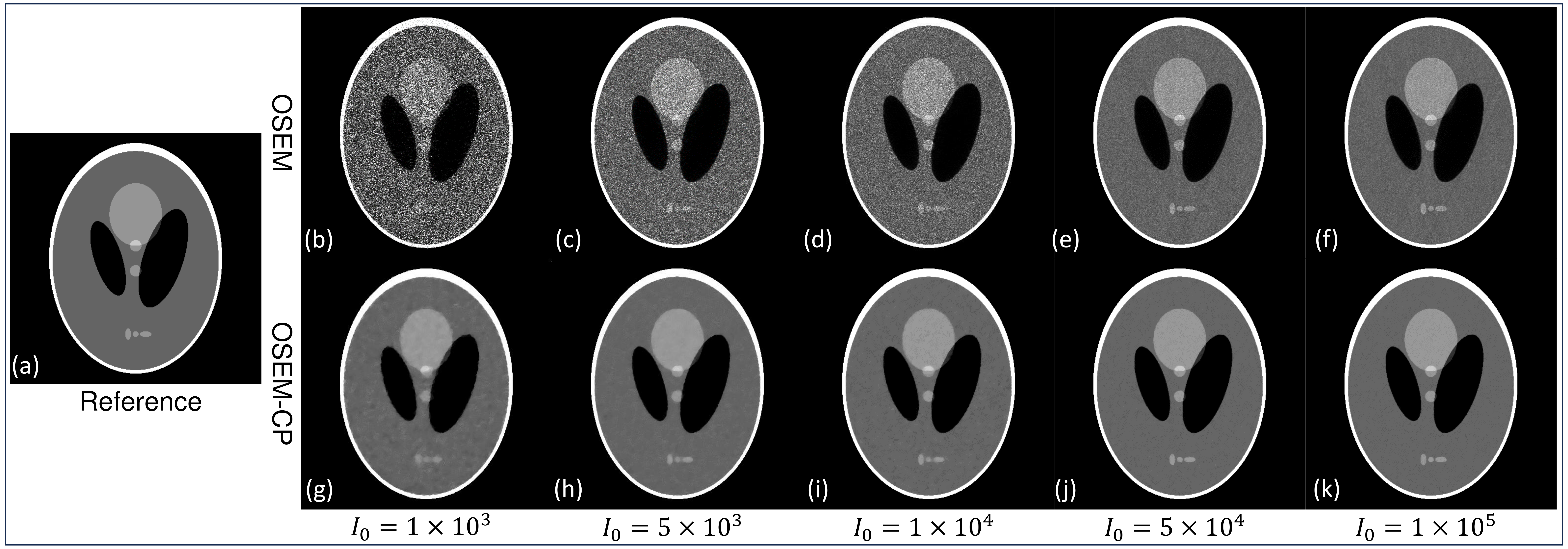}
  \vspace{-0.3cm}
	\caption{The \textit{Shepp-Logan} image reconstruction results of the OSEM algorithm and the proposed OSEM-CP at different dose levels from $I_0=1\times{10^3}$ to $I_0=1\times{10^5}$. The displayed windows of the gray values are $[0,0.5]$.}
	\label{fig_2_Robu_SL}
\end{figure*}

The proposed OSEM-CPNN contains $M$ sub-modules and $2M$ CNNs according to the number of ordered subsets. Each CNN contains 3 convolutional layers and 2 ReLU functions with a residual connection, the channel numbers of the middle layers are 32, and all the input variables of each CNN are integrated through a concatenation. The structure of the $n_{th}$ sub-modules is shown in Figure \ref{fig_1_net}. The black box is the optimization module of the dual problem where we used a CNN to optimize the proximal operator. The red box is the optimization module of the prime problem where we used another CNN here to learn the step size while retaining the form of the proximal operator.

\section{Experiments} 
To evaluate the performance of the proposed OSEM-CP algorithm and OSEM-CPNN, we performed experiments on both synthesized images and publicly available medical image sets. For the OSEM-CP, we performed robustness tests and competing tests against several popular non-learning methods. To verify the effectiveness and efficiency of OSEM-CPNN, we performed experiments on two datasets against several state-of-the-art supervised deep-learning methods.

\subsection{Experiments with the OSEM-CP algorithm}

Two experiments are performed to verify the effectiveness and performance, respectively, of the proposed OSEM-CP algorithm. The clean projection data were generated under a fan-beam imaging system with $720$ projection angles uniformly distributed in the range $[0, 2\pi]$ and a linear detector with $1024$ cells.

\subsubsection{Effectiveness verification}
We first employed the \textit{Shepp-Logan} phantom (as shown in Figure \ref{fig_2_Robu_SL}.(a)), with the gray value range of $[0, 1]$ and size of $512 \times 512$, to test the effectiveness of the proposed OSEM-CP algorithm, especially to verify if the strategy of integrating the regularization to the ``M''-step works as expected. To simulate the projection data at different doses, we added Poisson noise to the raw data, which is given by:
\begin{equation}
 \mathbf{p}_{n} = -\ln(\frac{I_{d}}{I_{0}}), I_{d} \sim \mathrm{Poisson}\{I_{0}\times e^{-\mathbf{p}_{c}}\},
	\label{eq_4.31}
\end{equation}
where $\mathbf{p}_{c}$ is the clean projection data and $\mathbf{p}_{n}$ the noisy one. The symbol $I_{0}$ denotes the number of incident photons and $I_{d}$ is the number of collected photons. Typically, a smaller $I_{0}$ means a lower dose and a higher noise level in the projection data. To test the robustness of OSEM-CP with regard to different noise levels, we set $I_{0}$ to be $1 \times {10^5}$, $5 \times {10^4}$, $1 \times {10^4}$, $5 \times {10^3}$ and $1 \times {10^3}$, respectively. 

The projection and reconstruction algorithms were coded by the \textbf{ODL} Python library \url{https://odlgroup.github.io/odl/index.html}. In terms of the evaluation indicators, we selected the commonly used PSNR and SSIM \cite{PSNR_hore2010image}. The reconstruction with the OSEM algorithm is employed as a reference. 

\begin{table}
	\centering
	\caption{The evaluation indicators of the results at different low-dose levels.}
	\begin{tabular}{ccccc} 
		\hline 
 \textbf{Method} &\multicolumn{2}{c}{\textbf{OSEM}} &\multicolumn{2}{c}{\textbf{OSEM-CP}} \\
		\textbf{Dose} ($I_{0}$)&\textbf{PSNR} &\textbf{SSIM} &\textbf{PSNR} &\textbf{SSIM} \\ 
 \cline{1-5} \\
 $1\times{10^3}$&16.01 &0.578 &\textbf{29.95} &\textbf{0.967} \\
		$5\times{10^3}$&22.94 &0.647 &\textbf{33.44} &\textbf{0.983} \\
		$1\times{10^4}$&25.84 &0.687 &\textbf{35.76} &\textbf{0.986} \\ 
		$5\times{10^4}$&31.72 &0.813 &\textbf{38.99} &\textbf{0.993} \\
 $1\times{10^5}$&33.72 &0.864 &\textbf{40.04} &\textbf{0.994} \\
		\hline
	\end{tabular}
	\label{table_1_robu} 
\end{table}

The reconstructions with the OSEM and OSEM-CP at different dose levels are shown in Figure \ref{fig_2_Robu_SL}. Compared with OSEM, OSEM-CP shows significant improvement in the image quality at each dose level. At higher doses, such as $I_0 = 1\times{10^5}$ and $I_0 = 5\times{10^4}$, the reconstructed images with OSEM-CP are very close to the reference ground-truth image. At the quite low-dose level, $I_0 = 1\times{10^3}$, noise in the result of OSEM submerges the circle and ellipsoid structures in the middle bottom region, while OSEM-CP successfully reconstructs these structures and separates the three ellipsoids clearly. Table \ref{table_1_robu} shows the quantitative measures PSNR and SSIM, where OSEM-CP exceeds OSEM by 6.32-13.94db in PSNR and 0.130-0.389 in SSIM, and the advantage is more significant at lower doses. The results show that the proposed OSEM-CP algorithm can conduct high-performance LDCT reconstructions and is robust against different dose levels.

\subsubsection{Performance evaluation}
\begin{figure}[!t]
	\centering
	\includegraphics[width=3.5in]{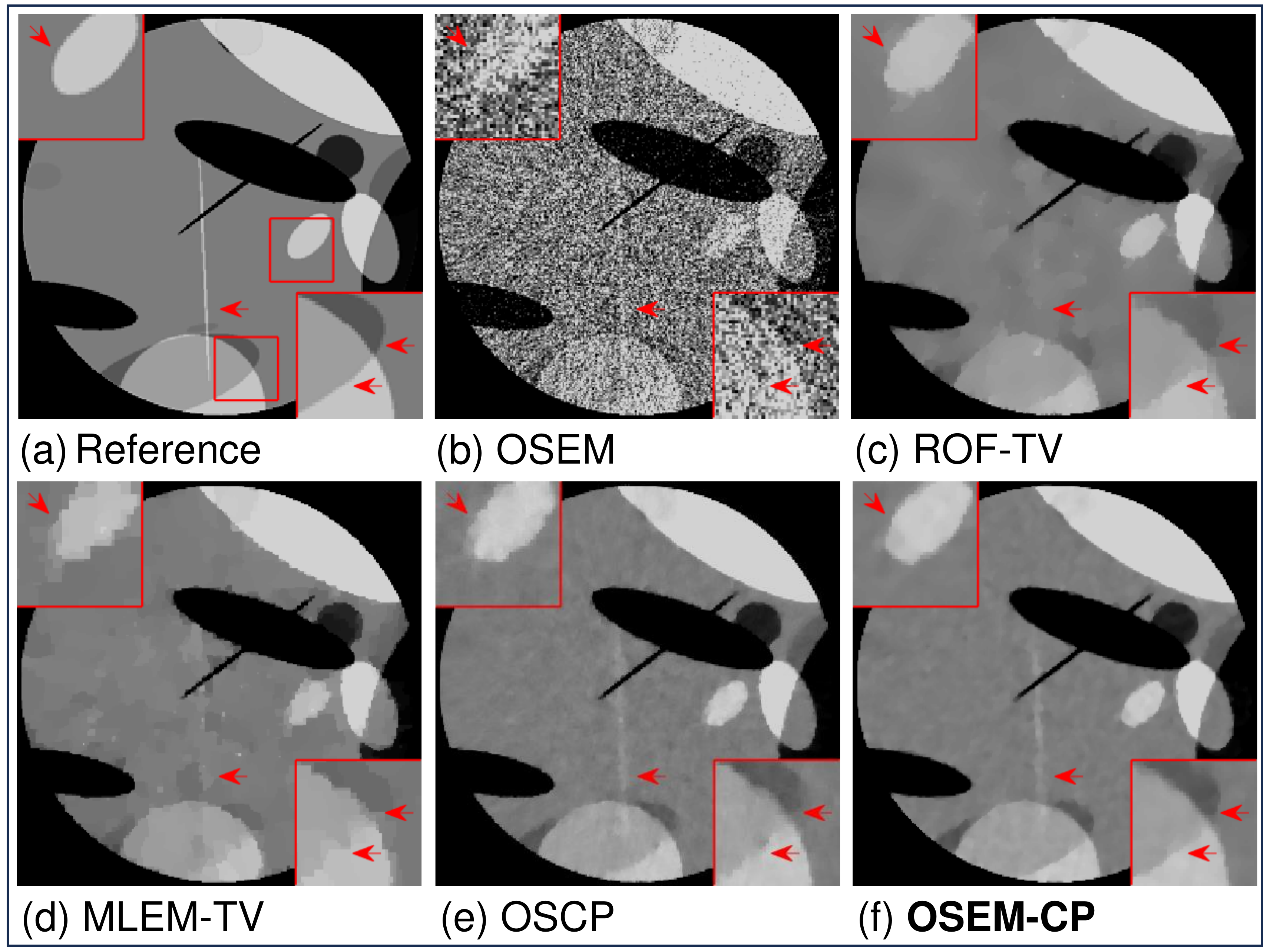}
	\vspace{-0.6cm}
	\caption{Results of the experiments with the \textit{Ellipses-010215} phantom at the low-dose conditions of $I_{0}=1\times{10^4}$. The displayed windows of the gray values are set to $[0.78, 1.45]$.}
	\label{fig_4_emcpEL}
\end{figure}

\begin{figure}[!t]
	\centering
	\includegraphics[width=3.5in]{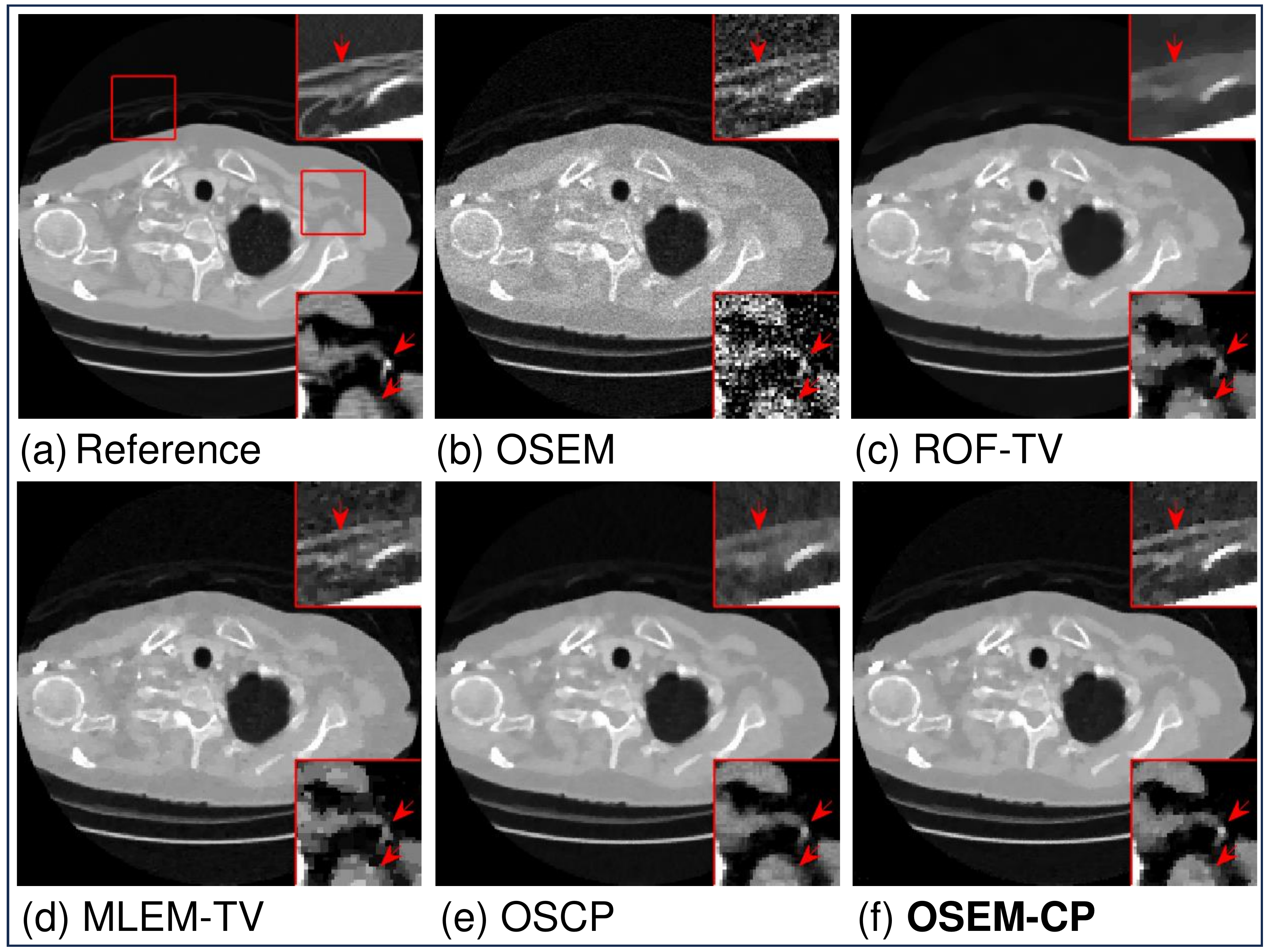}
	\vspace{-0.6cm}
	\caption{Results of the experiments with the \textit{IDRI-0001-012} phantom at the low-dose conditions of $I_{0}=5\times{10^4}$. The display windows are set to $[-1024, 2048]$ HU globally, $[-1024, -410]$ HU in the upper zoomed-in areas and $[783, 1295]$ HU in the lower zoomed-in areas.}
	\label{fig_5_emcpIDRI}
\end{figure}

To further verify the performance of OSEM-CP, we conducted experiments on three different phantoms against popular classic non-learning methods. In addition to the \textit{Shepp-Logan} phantom, we employed two more complex ones. One is the \textit{Ellipses-010215} phantom generated by the \textbf{ODL} library, which contains several overlapping ellipses, with gray value range $[0,1]$ and size of $256 \times 256$, as shown in Figure \ref{fig_4_emcpEL}(a). The other one is a medical image \textit{IDRI-0001-012} chosen from the \textit{LIDC-IDRI} dataset, with size of $256 \times 256$ and HU range of $[-1024, 2048]$, as shown in Figure \ref{fig_5_emcpIDRI}(a). The experimental setup for the \textit{Shepp-Logan} phantom is the same as that in the previous experiment. For the phantoms \textit{Ellipses-010215} and \textit{IDRI-0001-012}, we used the CT system with $360$ projection angles and a $512$-cell detector. The low-dose projection data were generated in the same way as the previous experiments, and the dose for the three phantoms are set to $5\times{10^3}$, $1\times{10^4}$ and $5\times{10^4}$, respectively. For reasonable comparison, we chose TV-regularized post-processing method ROF-TV \cite{ROF_rudin1992nonlinear}, and two TV-regularized reconstruction methods OSCP \cite{OSCP_deng2019fast} and MLEM-TV \cite{friot2022iterative}, as the competing methods. ROF-TV utilizes a TV-regularization term to post-process the noisy image according to the patch-constant property of the ideal clean image. OSCP imposes TV-regularization constraint in the iterations of OS-SART and solves with the CP algorithm. MLEM-TV introduces the TV-regularization term into the MLEM algorithm and uses a two-step alternating iteration approach to perform reconstruction and regularization. The iteration numbers and hyper-parameters of all the methods are well-adjusted to achieve the best indicators (PSNR and SSIM) on the shown phantoms.

\begin{figure}[!t]
	\centering
	\includegraphics[width=3.5in]{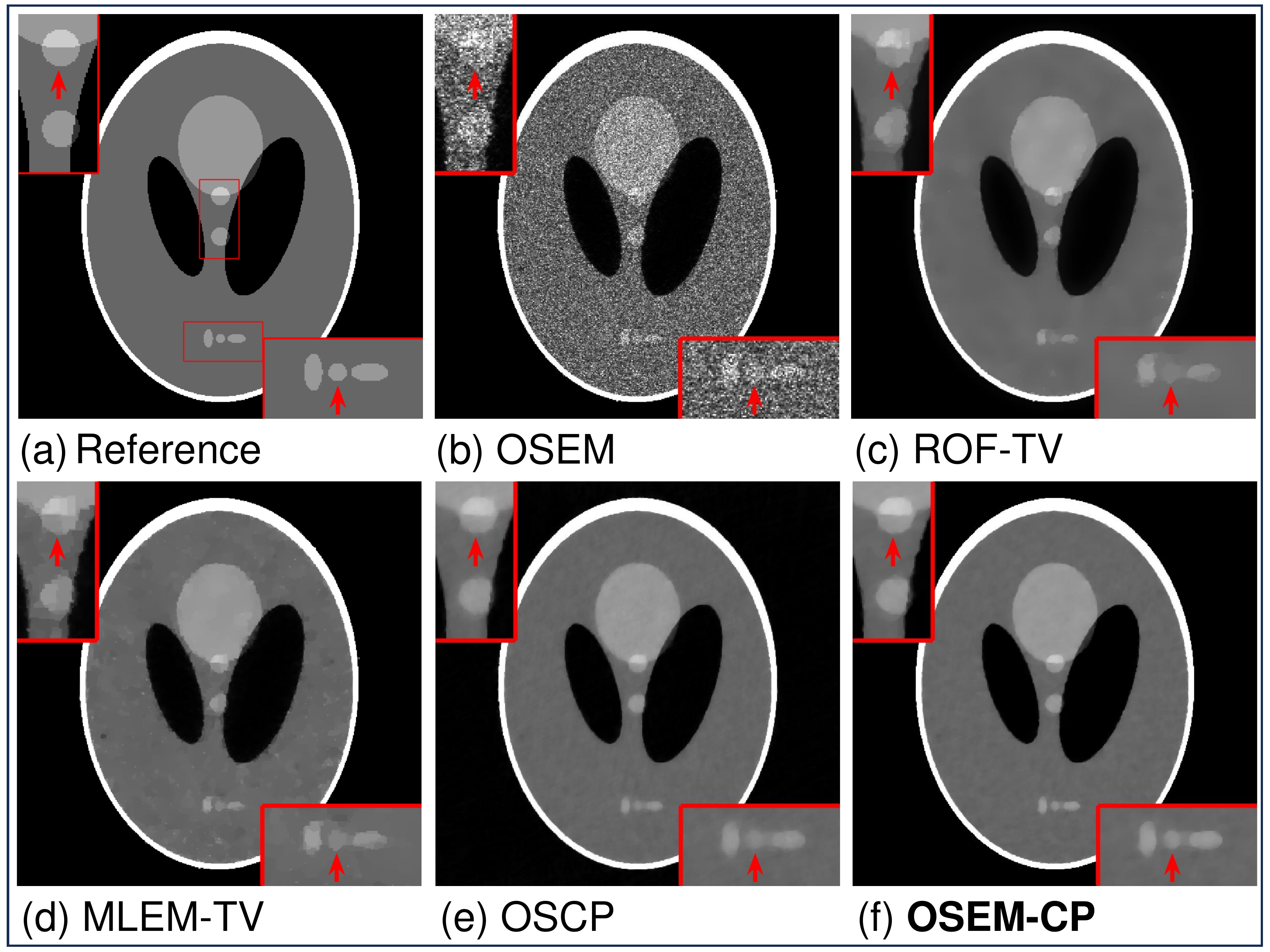}
	\vspace{-0.6cm}
	\caption{Results of the experiments with the \textit{Shepp-Logan} phantom at the low-dose conditions of $I_{0}=5\times{10^3}$. The displayed windows of the gray values are set to $[0,0.5]$.}
	\label{fig_3_emcpSL}
\end{figure}

\begin{table}[!t]
	\centering
	\caption{The evaluation indicators for the results of the experiments on the three phantoms.}
	\begin{tabular}{lcccccc} 
		\hline 
 \textbf{Model} &\multicolumn{2}{c}{\textit{Shepp-Logan}} &\multicolumn{2}{c}{\textit{Ellipses-010215}} &\multicolumn{2}{c}{\textit{IDRI-0001-012}} \\
		\textbf{Method}&\textbf{PSNR} &\textbf{SSIM} &\textbf{PSNR} &\textbf{SSIM} &\textbf{PSNR} &\textbf{SSIM}\\ 
 \cline{1-7}\\
		\textbf{OSEM} &22.94 &0.647 &21.71 &0.391 &31.79 &0.825 \\
		\textbf{ROF-TV} &31.01 &0.922 &33.29 &0.931 &36.39 &0.947 \\
 \textbf{MLEM-TV} &30.46 &0.959 &33.85 &0.951 &36.13 &0.940 \\
 \textbf{OSCP} &31.59 &0.921 &35.06 &0.948 &36.92 &0.950 \\
 \textbf{OSEM-CP} &\textbf{33.44} &\textbf{0.983} &\textbf{36.12} &\textbf{0.967} &\textbf{37.57} &\textbf{0.958} \\
		\hline
	\end{tabular}
	\label{table_2_emcp} 
\end{table}

\begin{figure*}[!t]
	\centering
	\includegraphics[width=7.1in]{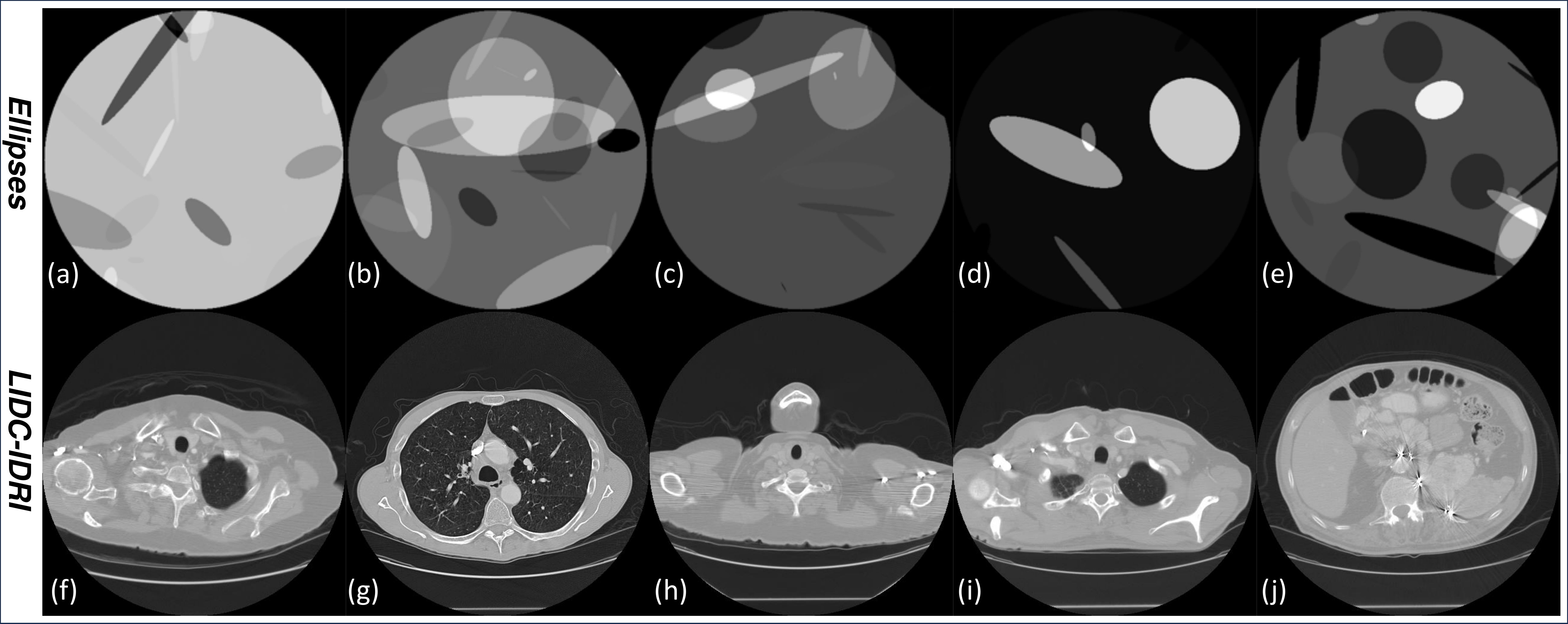}
	\caption{Typical examples of the \textit{``Ellipses''} dataset ((a)-(e)) and the \textit{``LIDC-IDRI''} dataset ((f)-(j)).}
	\label{fig_6_exs}
\end{figure*}

Figure \ref{fig_3_emcpSL}, \ref{fig_4_emcpEL} and \ref{fig_5_emcpIDRI} show the results of the experiments with the three phantoms, respectively. OSEM-CP achieved outstanding reconstructions on all three phantoms and each low-dose level. As shown in the zoomed-in areas of Figure \ref{fig_3_emcpSL}, OSEM-CP restores the circles and ellipsoids in the \textit{Shepp-Logan} phantom with the highest sharpness and most accurate edges compared with other methods. In Figure \ref{fig_4_emcpEL}, the \textit{Ellipses-010215} phantom has bad contrast and many elliptical arcs. However, OSEM-CP still works well on the shape and edge smoothness of these parts, as shown in the upper zoomed-in area, and offers better contrast and accuracy in the low-contrast overlapping arcs, as shown in the lower zoomed-in area. Also, OSEM-CP performs well in reconstructing the longitudinal and elongated ellipsoid in the centre of the image, with the best effects in its sharpness and completeness, as indicated by the arrow. In Figure \ref{fig_5_emcpIDRI}, OSEM-CP provides better structure accuracy and sharpness while effectively suppressing noise. The upper zoomed-in area shows that OSEM-CP ensures denoising effects without blurring. In the lower zoomed-in area, OSEM-CP offers the best sharpness and contrast at the dot that is pointed by the upper arrow and the arc edge that is indicated by the lower arrow. Table \ref{table_2_emcp} shows the evaluation indicators for the results with those three phantoms. Clearly, OSEM-CP has significant advantages over competing methods. 

The performance advantage of the proposed OSEM-CP algorithm over competing methods is based on several aspects. As a pure post-processing method, ROF-TV cannot participate in reconstruction, thus easily causing structural deformation. Although the MLEM-TV algorithm integrates regularization into its iterations, its two-step alternating strategy is similar to progressive post-processing and the TV-regularization term does not really constrain the reconstruction. In addition, the OS strategy for optimizing the iteration is also an advantage of our OSEM-CP over the MLEM-TV. The OSCP algorithm, which also uses the OS strategy and CP algorithm to solve the TV-regularization problem, has good results, but lacking statistical priors still makes it inferior to our OSEM-CP. In contrast, our OSEM-CP directly solves the TV-regularized EM model by the CP algorithm and OS strategy, which not only incorporates the statistical priors into LDCT reconstruction effectively but also makes up for the respective drawbacks of the above methods, therefore achieving better results.

\subsection{Experiments with OSEM-CPNN}
Our experiments for the proposed OSEM-CPNN were performed on two datasets: $(1)$ the \textit{Ellipses} dataset \cite{learnpd_adler2018learned} and $(2)$ the \textit{LIDC-IDRI} dataset \cite{idri_armato2011lung}. The \textit{Ellipses} dataset is a simulated-image dataset generated by the \textbf{ODL} library, containing images occupied by overlapping ellipses. Several example images are shown in the first row of Figure \ref{fig_6_exs}. The \textit{LIDC-IDRI} dataset is a medical-image dataset that consists of $241689$ normal-dose CT image slices of $1012$ different patients, and several examples are shown in the second row of Figure \ref{fig_6_exs}. For both datasets, we randomly chose 256 images as the training set, 128 for validation, and 20 for testing. We interpolated all the images to a size of $256 \times 256$ and generated the low-dose projection data using a simulated imaging system with 360 projection angles and a detector with 512 cells, following the same way as the previous experiments. The dose was set as $I_{0}=1\times{10^4}$ for experiments on both datasets.

In terms of the comparative methods, apart from the traditional ones in the previous experiments for OSEM-CP, we additionally chose the non-local mean filter NLM \cite{NLM_buades2005non} and three popular state-of-the-art supervised deep learning methods: REDCNN \cite{REDCNN_chen2017low}, CTFormer \cite{wang2023ctformer}, and DRCNN \cite{drcnn_feng2021dual}. REDCNN employs a residual encoder-decoder network to perform LDCT image denoising, CTFormer uses the transformer structures instead of the CNNs to denoise LDCT images, and DRCNN is a dual-domain method that performs supervised denoising in both the sinogram and image domains, with a TV-regularization term in its loss function. Codes for all the comparative methods are publicly available, and all the CT reconstructions are conducted based on the \textbf{ODL} library. We used the MSE as the loss function and the Adam \cite{ADAM_kingma2014adam} algorithm with an initial learning rate of 0.0001 for optimization. Limited by storage, the mini-batch size of OSEM-CPNN was set to 1, while that of the other deep learning methods was 8. All experiments were performed on a server running Ubuntu 20.04.5 with Python 3.7.16, PyTorch 1.13.11, and a Nvidia Tesla V100 GPU card.

\begin{figure*}[!t]
	\centering
	\includegraphics[width=7.1in]{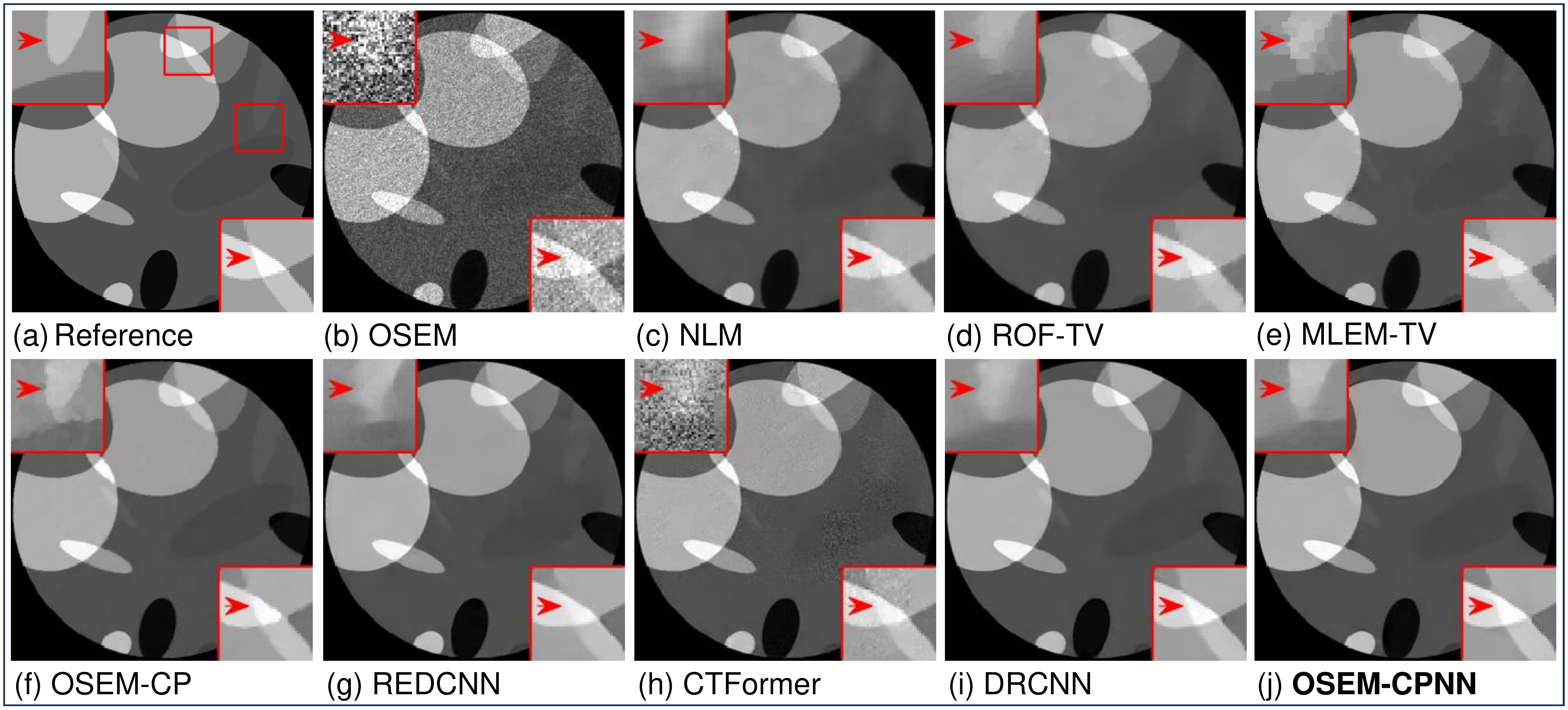}
	\vspace{-0.3cm}
	\caption{Results of the experiments with the \textit{Ellipses} dataset at the low-dose conditions of $I_{0}=1\times{10^4}$. The display windows of the gray values are set to $[0,1.0]$ globally and $[0, 0.4]$ in the upper zoomed-in areas.}
	\label{fig_7_emcpnnEL}
\end{figure*}

\begin{figure*}[!t]
	\centering
	\includegraphics[width=7.1in]{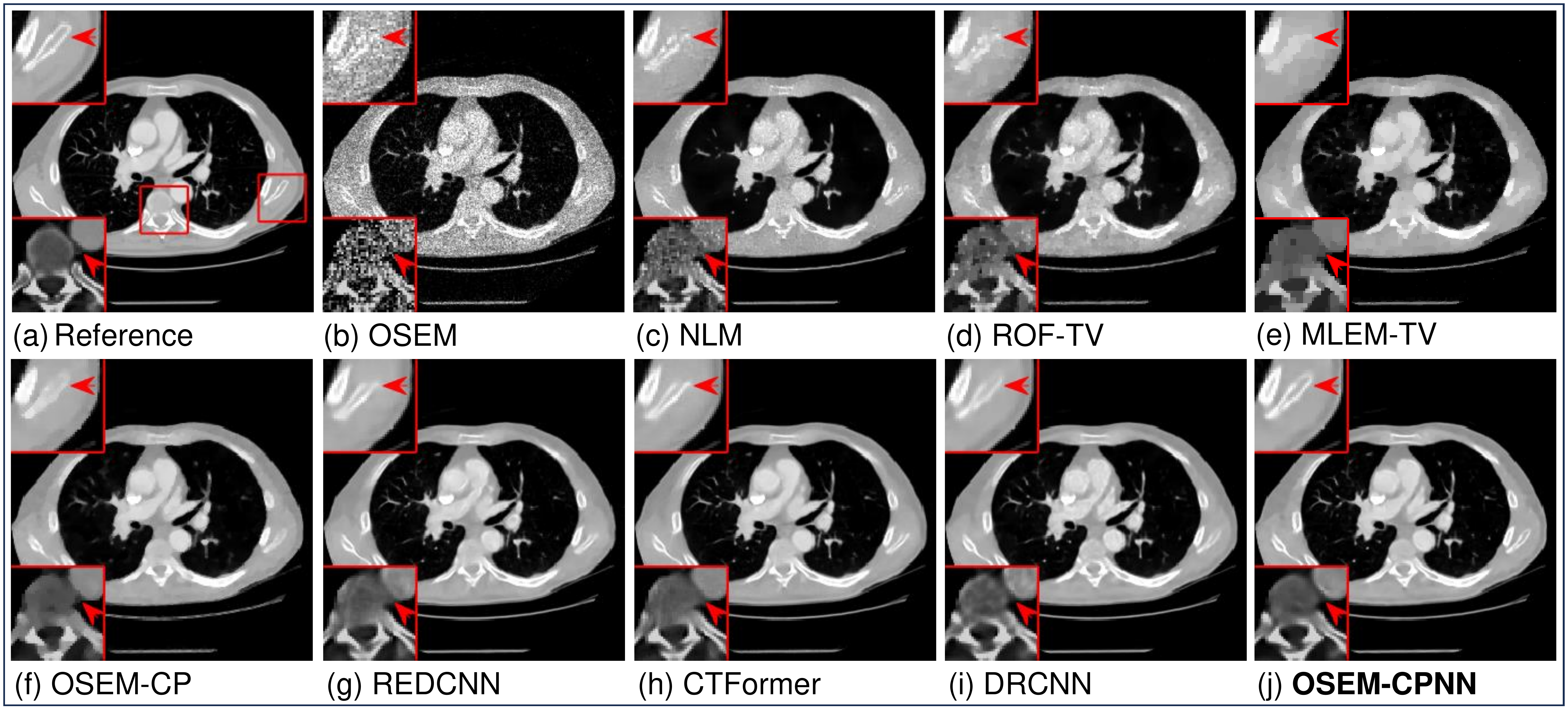}
	\vspace{-0.3cm}
	\caption{Results of the experiments with the \textit{LIDC-IDRI} dataset at the low-dose conditions of $I_{0}=1\times{10^4}$. The display windows are $[-420,1620]$ globally and $[780, 1800]$ in the lower zoomed-in areas.}
	\label{fig_8_emcpnnIDRI}
\end{figure*}

Figure \ref{fig_7_emcpnnEL} shows the results of the experiments on the \textit{Ellipses} dataset at dose $I_{0}=1\times{10^4}$. OSEM-CPNN demonstrates the best reconstruction quality among the competing methods. As shown in the upper zoomed-in area, the ellipsoid pointed by the arrow has the closest shape and edges with the best sharpness compared to other methods. In the lower zoomed-in area, OSEM-CPNN reconstructs the clearest intersecting lines in the overlapping areas of multiple ellipses and provides the best contrast. Figure \ref{fig_8_emcpnnIDRI} shows the results of the experiments on the \textit{LIDC-IDRI} dataset at dose $I_{0}=1\times{10^4}$. As shown in the upper zoomed-in area, OSEM-CPNN presents the best shape and completeness on the oval pointed by the arrow. In the lower zoomed-in area, OSEM-CPNN shows effective denoising while clearly separating the structures where other methods exhibit adhesion, as indicated by the arrows. Table \ref{table_3_emcpnn} lists the average evaluation indicators (psnr(db)/ssim) of each method on the test sets. In the experiments on the \textit{Ellipses} dataset, OSEM-CPNN leads ahead by 0.88db in terms of PSNR and 0.006 with regard to SSIM. For the experiments with the \textit{LIDC-IDRI} dataset, the leading amounts are 0.3db and 0.005, respectively. Both the visual results and the evaluation indicators demonstrate the outstanding performance of the proposed OSEM-CPNN.

\begin{table}[!t]
	\centering
	\caption{The average indicators (psnr(db)/ssim), average inference time (seconds), and parameter number (millions) of each method on the test sets of both datasets.}
	\begin{tabular}{lcccccc} 
		\hline 
 \textbf{Dataset} &\multicolumn{2}{c}{\textit{Ellipses}} &\multicolumn{2}{c}{\textit{LIDC-IDRI}} \\
		\textbf{Method}&\textbf{PSNR} &\textbf{SSIM} &\textbf{PSNR} &\textbf{SSIM} &\textbf{Time} &\textbf{Para}\\ 
 \cline{1-7} \\
		\textbf{OSEM} &24.30 &0.463 &25.57 &0.630 &1.25s &- \\
		\textbf{NLM} &35.52 &0.921 &31.80 &0.866 &3.16s &- \\
 \textbf{ROF-TV} &35.93 &0.947 &31.98 &0.889 &1.32s &- \\
 \textbf{MLEM-TV} &36.03 &0.958 &31.73 &0.884 &19.8s &- \\
		\textbf{OSEM-CP} &39.12 &0.978 &33.07 &0.925 &71.9s &- \\
 \textbf{REDCNN} &39.44 &0.979 &34.58 &0.930 &1.33s &8.02M \\
 \textbf{CTFormer} &35.09 &0.863 &35.23 &0.938 &1.44s &9.13M\\
 \textbf{DRCNN} &39.86 &0.963 &36.04 &0.942 &0.22s &8.32M\\
 \textbf{OSEM-CPNN} &\textbf{40.86} &\textbf{0.985} &\textbf{36.49} &\textbf{0.950} &3.63s &7.20M\\
		\hline
	\end{tabular}
	\label{table_3_emcpnn} 
\end{table}

Compared with other deep learning methods, the proposed OSEM-CPNN requires longer training and inference time with a 
similar parameter size. Fortunately, its inference time is still at a fast level, which ensures its practical availability, as shown in Table \ref{table_3_emcpnn}. At the same time, OSEM-CPNN has quite apparent advantages. Learning the proximal operators and hyperparameters instead of a complete denoiser makes it suitable for fine-tuning over datasets, which introduces high practicality. In our experiments, the network of the experiments on the \textit{LIDC-IDRI} dataset was fine-tuned for $50$ epochs from the trained network on the \textit{Ellipses} dataset, which is entirely different in the image type. In contrast, complete training requires more than $1000$ epochs, which reflects the excellent fine-tuning adaptability of the proposed OSEM-CPNN. Furthermore, the outstanding performance of OSEM-CPNN makes it competitive with the state-of-the-art deep-learning methods or even shows superiority.

\section{Discussion and Conclusion}
We proposed a statistical iterative algorithm OSEM-CP and an end-to-end network OSEM-CPNN for low-dose CT reconstruction. By directly solving the TV-regularized EM minimization problem with the CP algorithm and employing the OS iteration strategy, OSEM-CP effectively integrates the statistical priors into the reconstruction process and enables high-quality and efficient LDCT reconstruction. As the unrolling network version of OSEM-CP, OSEM-CPNN utilizes neural networks to replace the hyperparameters and proximal operators in the OSEM-CP and learns them implicitly and adaptively. While eliminating the workload for manual parameter tuning, OSEM-CPNN greatly reduces the computation for reconstruction to only one full-view iteration and significantly improves the reconstruction performance. At the same time, OSEM-CPNN has good fine-tuning adaptation which is valuable in actual applications. Experiments on different types of image models verified the promising reconstruction effect of the proposed OSEM-CP algorithm and its robustness to various low-dose noise levels. Results on two datasets containing different types of images also demonstrated that the proposed OSEM-CPNN has exceptional performance rivaling or even surpassing some state-of-the-art deep-learning methods.

Although using the emission EM algorithm is imprecise, it still provides more priors than methods that do not consider statistical priors, especially in very low-dose levels. With the TV regularization, the negative effect of model bias can be greatly reduced, and the statistical priors would then promote a better reconstruction, which can be certified by comparing the OSCP and the proposed OSEM-CP. We believe that the proposed OSEM-CP will be a good choice for regularized iterative LDCT reconstruction, especially under highly low-dose conditions. On the other hand, the proposed OSEM-CPNN is a high-performance deep-learning option when training data is available. Compared with post-processing methods, the inference time of OSEM-CPNN is longer. However, it is still at a fast and practicable level and much faster than traditional iterative ones.

We have shown that statistical priors would improve LDCT reconstructions. To statistically model the projection data better, the transmission EM algorithm should be employed, instead of the emission EM algorithm. The main obstacle is how to solve the ``M''-step accurately and efficiently. We will study this in our future work.

\IEEEpeerreviewmaketitle

\section*{Acknowledgment}
This work was supported by Beijing Natural Science Foundation (No.Z210003), National Natural Science Foundation of China (NSFC) (61971292) and China Scholarship Council (CSC) (No.202307300001). The authors are also grateful to Beijing Higher Institution Engineering Research Center of Testing and Imaging for funding this research work.

\ifCLASSOPTIONcaptionsoff
 \newpage
\fi

\bibliography{sec_emcpnn}
\bibliographystyle{IEEEtran}

\end{document}